# Multi-Stage Feature Selection Based Intelligent Classifier for Classification of Incipient Stage Fire in Building

**Allan Melvin Andrew \*, Ammar Zakaria, Shaharil Mad Saad and Ali Yeon Md Shakaff**

Centre of Excellence for Advanced Sensor Technology (CEASTech), Universiti Malaysia Perlis, Jejawi, 02600 Arau, Perlis, Malaysia; sag.unimap@gmail.com (A.Z.); shaharil85@gmail.com (S.M.S.); aliyeon@unimap.edu.my (A.Y.M.)

\* Correspondence: allanmelvin@unimap.edu.my; Tel.: +60-174299713

**Abstract:** In this study, an early fire detection algorithm has been proposed based on low cost array sensing system, utilising off- the shelf gas sensors, dust particles and ambient sensors such as temperature and humidity sensor. The odour or "smellprint" emanated from various fire sources and building construction materials at early stage are measured. For this purpose, odour profile data from five common fire sources and three common building construction materials were used to develop the classification model. Normalised feature extractions of the smell print data were performed before subjected to prediction classifier. These features represent the odour signals in the time domain. The obtained features undergo the proposed multi-stage feature selection technique and lastly, further reduced by Principal Component Analysis (PCA), a dimension reduction technique. The hybrid PCA-PNN based approach has been applied on different datasets from in-house developed system and the portable electronic nose unit. Experimental classification results show that the dimension reduction process performed by PCA has improved the classification accuracy and provided high reliability, regardless of ambient temperature and humidity variation, baseline sensor drift, the different gas concentration level and exposure towards different heating temperature range.

## 1. Introduction

Fires can be categorized into two main groups: direct burning and indirect burning. Residential fires may happen indoors or outdoors [1]. Most fires start from an incipient stage and develop further to smouldering, flaming and fire stages [2]. In incipient and smouldering cases, fires have less flames and smoke, while in the flaming and fire stages, fires have more flames and radiate extreme heat.

According to the work published in the recent decade, fire research can be categorized mainly into four types; namely, fire detection, fire prediction, fire data analysis and reduction of false fire alarms [2]. Predicting or perceiving fire at the early stage is very challenging and crucial for both personal and commercial applications. Over the years, several methods have been proposed which utilise various sensing technologies to provide early fire detection [2]. The research conducted by Rose-Pehrsson is able to provide early fire detection using a Probabilistic Neural Network and achieves higher classification accuracy [3]. However, they were only able to demonstrate it as early as the smouldering stage. As for data analysis alone, various methodologies have been utilised. The most common methods used are related to clustering techniques and classification algorithms.

Several fire data analysis algorithms have been proposed. According to the research, most of these algorithms are based on time-fractal approaches to characterize the temporal distribution of detected fire sequences [4]. Some of the research has focused on utilizing unsupervised ways to detect fire from the signals [5]. In their paper, Chakraborty and Paul proposed a hybrid clustering algorithm using a modified k-means clustering algorithm. Although it required very little processing time and managed to detect the fire flames at fast speed, the proposed algorithm can be only be used in video image processing based on RGB and HSI colour models. Bahrepour *et al.*, in their research, investigated the feasibility of spatial analysis of indoor and outdoor fires using data mining approaches for WSN-based fire detection purposes [6]. In their paper, they had investigated the most dominant feature in fire detection applications. Kohonen self-organizing map (kSOM) had been utilized as a feature reduction technique which can cluster similar data together. Experimentals result show that their method reduces the number of features representing the fire data features. They also performed analysis on residential fires and used artificial neural network, naive Bayes and decision tree classifiers to compute the best combination of sensor type in fire detectors. The outputs of various classifiers were fused using data fusion techniques to achieve higher fire detection accuracy. The reported results showed that 81% accuracy for residential fire detection and 92% accuracy for wildlife fire detection could be achieved.

Most of the proposed methods provide high classification rates in detecting fires, albeit they need to be in close vicinity to the source of the fire and only operate based on specific types of sensors [7–13]. Mimicking the human nose in early fire detection is still the biggest challenge for olfactory engineering. The present electronic nose systems have difficulties in detecting early fires, especially in large spaces, and cannot provide additional information regarding the burning stages and the scorching fire material. To overcome the mentioned weakness, bio-inspired approaches based on electronic nose technology is a promising method, which utilises artificial intelligence in detecting and predicting the possibility of fire occurrence. Although there are many proposed feature selection techniques and classifiers involved, the real question is whether it is possible to implement them in conventional fire detectors, yet to be determined, at a low cost. This paper focuses on investigating a multi-stage feature selection method using a bio-inspired artificial neural network and principal component analysis for data reduction, which can give the best detection accuracy, reduce misclassification and offer high reliability for indoor fire detection applications. This work is important to investigate the most suitable features and classification algorithm, which could be proved less computationally complex and having potential to be used in embedded applications.

The rest of this paper is organized as follows: Section 2 introduces the features of fires. Section 3 describes the proposed four-stage fire detection algorithm. Section 4 discusses the experimental results of the proposed method and compares the performance of the proposed method with those of other fire detection algorithms, and Section 5 presents the conclusions of our study.

## 2. Methods

In this section, the odour measurement technique, the feature extraction from sensor arrays using various data normalisation techniques, the artificial neural network-based feature selection, the feature reduction using PCA, and the classification stages are explained. Figure 1 shows the flowchart of the proposed multi-stage feature selection approach using PCA and PNN. The dashed line around PNN training on training dataset in Figure 1 indicates that the PNN training is conducted prior to the classification of fire sources. The training dataset is used by PNN in the fire sources classification process.

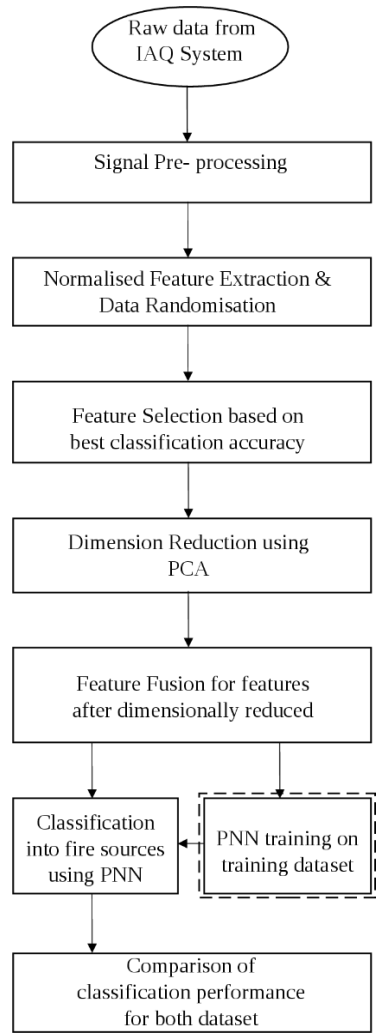

**Figure 1.** A flowchart of the proposed multi- stage feature selection approach using PCA and PNN.

*2.1. Datasets*

In this study, two datasets have been used. The first dataset consists of odour signals which have been obtained from an in-house metal oxide gas sensor-based low cost (IAQ) system, consisting of oxygen ($O_2$), volatile organic compound (VOC), carbon dioxide ($CO_2$), ozone ($O_3$), nitrogen dioxide ($NO_2$), particulate matter up to 10 micrometres in size ($PM_{10}$), temperature and humidity sensors. The prediction classifier for the early fire detection has been developed based on odours from various sample sources. The odour sources consist of five common fire sources and three common building construction materials. Information about the materials tested and their sample dimensions prepared according to the corresponding European Standard, is shown in Table 1. For each source, more than 100 odour measurement samples have been taken at seven different temperature points, starting from 50 °C up to 250 °C. About 200 ambient air measurement datapoints have been added to the dataset as a reference air sample. The ambient air samples are considered the 9th tested sample in this paper. The final IAQ system dataset is a matrix of 1000 rows and eight columns. The training set contains 600 samples (60% of the dataset), the validation set contains 100 samples (10% of the dataset), and the test set contains the remaining samples, which is 30% of the dataset. In order to estimate the true performance of the classifier, the test is based on the remaining samples which were not used during the training and validation process. The dataset has been referred as the IAQ dataset in this paper.

**Table 1.** The tested materials and its sample dimension prepared according to European Standard.

| Sample | Materials | Material Type | Dimension |
|--------|-----------|---------------|-----------|
| Sample 1 | Paper | Common Fire Source | 16 pieces 5 cm × 5 cm<br>90 gsm sheets stacked together |
| Sample 2 | Plastic | Common Fire Source | 4 cm × 2 cm × 40 cm (density 20 kg·m$^{-3}$)<br>polyurethane |
| Sample 3 | Styrofoam | Common Fire Source | 4 cm × 2 cm × 40 cm styrofoam |
| Sample 4 | Cotton | Common Fire Source | 1 wick 18 cm long (approx. 0.17 g) |
| Sample 5 | Cardboard | Common Fire Source | 16 pieces 5 cm × 5 cm stacked together |
| Sample 6 | Wood | Building Construction Material | 1 cm × 1 cm × 2 cm beech wood |
| Sample 7 | Brick | Building Construction Material | 1 piece brick |
| Sample 8 | Gypsum board | Building Construction Material | 1 cm × 1 cm × 2 cm gypsum board |

The second dataset obtained from a Portable Electronic Nose (PEN3) from Airsense Analytics GmbH (Schwerin, Germany) has been used as the control dataset. This set has 10 sensor inputs (10 columns). For each source, more than 100 samples of odour measurements have been taken at seven temperature points, starting from 50 °C up to 250 °C. Like IAQ, 200 ambient air measurement datapoints have been added to the dataset as a reference air sample. The final PEN3 dataset is a matrix of 1000 rows and 10 columns. The training set contains 600 samples (60% of the dataset), the validation set contains 100 samples (10% of the dataset), and the test set contains the remaining samples, which is 30% of the dataset, similar to the first dataset. A similar approach for performance analysis was followed for the above process as with IAQ. The dataset is referred to as PEN3 dataset in this paper.

*2.2. Measurement of Odour Signals*

In the IAQ dataset, the odour samples have been collected from the IAQ system placed at 2.1 m height in the testing room. The height of 2.1 m has been selected to deploy the in-house system in buildings based on few classification preliminary tests done at different heights in a standard sized room (33 m³ in volume) in Malaysia. Heights of 0.7, 1.4 and 2.1 m have been tested in the preliminary tests. A height of 2.1 m was the most suitable and was been selected because the experimental results show that the gases generated at the incipient fire stage fill the top part of the room first since the density of the emitted gases are lesser than that of ambient air. For this experiment, the deployment of the sensor unit at this height gives the best chance in predicting an earlier fire event. Having the sensor units deployed at an inappropriate height in the building can cause it to miss useful data for fire data analysis and prediction, and thus, could trigger false fire alarms. That is also the main reason why conventional fire detectors are placed on the ceilings of buildings [14]. For realisation of a wireless sensing IAQ system, the data of the low cost system is sampled at the sampling rate of 10 sample/min [15]. The data has been recorded for 15 min each time. Each data measurement has been sent wirelessly to the server for processing and data storage using an available wireless sensor network. The data measurements have been recorded in websocket "*sqlite*" format and then converted to ".csv" format using a custom LabVIEW application. Afterwards, the odour signals have been translated into digital form by a custom MATLAB application.

In the PEN3 dataset, the data from PEN3 has been captured using a program supplied by AirSense Analytics GmbH. The PEN3 has been placed at 1.5 m distance from the smell source which has been heated in a vacuum oven. PEN3 has a sampling frequency of 1 sample/s. The data has been recorded for 15 min each. The data measurements have been recorded in ".nos" format and then converted to ".xls" format using a custom application. Then, the samples have been converted into digital format by a custom MATLAB application.

*2.3. Normalised Feature Extraction*

Baseline drift is a widespread phenomenon in signal analysis, which could also cause incorrect representation of data in subsequent feature extraction and feature selection processes of an odour signal, and baseline correction is the solution to the problem and the correct way of representing the signal when the analysis deals with sensor values from different conversion units. Baseline manipulation helps to pre-process the sensor output to free itself from the drift effect, the intensity dependence and, possibly, from non-linearity [7].

In this paper, for the feature extraction stage, five types of baseline correction algorithms have been executed on both datasets by converting the raw data value from Volts to unit ratio values. Unit ratio is a dimensionless unit. Each type of baseline correction has been considered as a feature. The ability to distinguish the fire event from the normalised data itself helps to reduce the computation complexity and classification time, thus it will be easier to implement it in the embedded system using C programming.

The first feature is Relative Logarithmic Sum Squared Voltage value (RLSSV). RLSSV is the division of logarithmic voltage by the logarithmic sum squared voltage value. The equation for calculating RLSSV is shown in Equation (1):

$$\text{RLSSV} = \frac{\log v_i}{\log(\sum v^2)} \qquad (1)$$

where $v_i$ is the voltage value at time *i* for each specific sensor.

The second feature is the Relative Logarithmic Voltage value (RLV). RLV is the ratio between the logarithmic voltage and the instantaneous voltage value. It can be calculated using Equation (2):

$$\text{RLV} = \frac{\log v_i}{v} \qquad (2)$$

where $v_i$ is the voltage value at time *i* for each specific sensor.

The next feature is Relative Sum Squared Voltage value, referred to as RSSV. RSSV is obtained by dividing the instantaneous voltage value by the square root value of sum of squared voltages. Equation (3) shows the formula used in computing the RSSV:

$$\text{RSSV} = \frac{v_i}{\sqrt{\sum v^2}} \qquad (3)$$

where $v_i$ is the voltage value at time *i* for each specific sensor.

The fourth feature is Relative Voltage value (RV). RV is calculated by finding the ratio of the voltage at time I and the average. It can be calculated using Equation (4):

$$\text{RV} = \frac{v_i}{v_o} \qquad (4)$$

where $v_i$ is the voltage value at time *i* and $v_0$ is the baseline voltage value for each specific sensor.

The final feature investigated is the Fractional Voltage Change value (FVC). FVC is directly proportional to the difference between the averaged baseline value and current value and indirectly proportional to the averaged baseline value, as shown in Equation (5):

$$\text{FVC} = \frac{\bar{v}_0 - v_i}{\bar{v}_0} \qquad (5)$$

where $v_i$ is the actual sensor value at time *i* and $\overline{v_0}$ is the baseline value of each specific sensor.

A raw data example of the scorching smell generated by paper at 250 °C and its waveform after the RLSSV feature has been extracted are presented in Figure 2a,b, respectively.

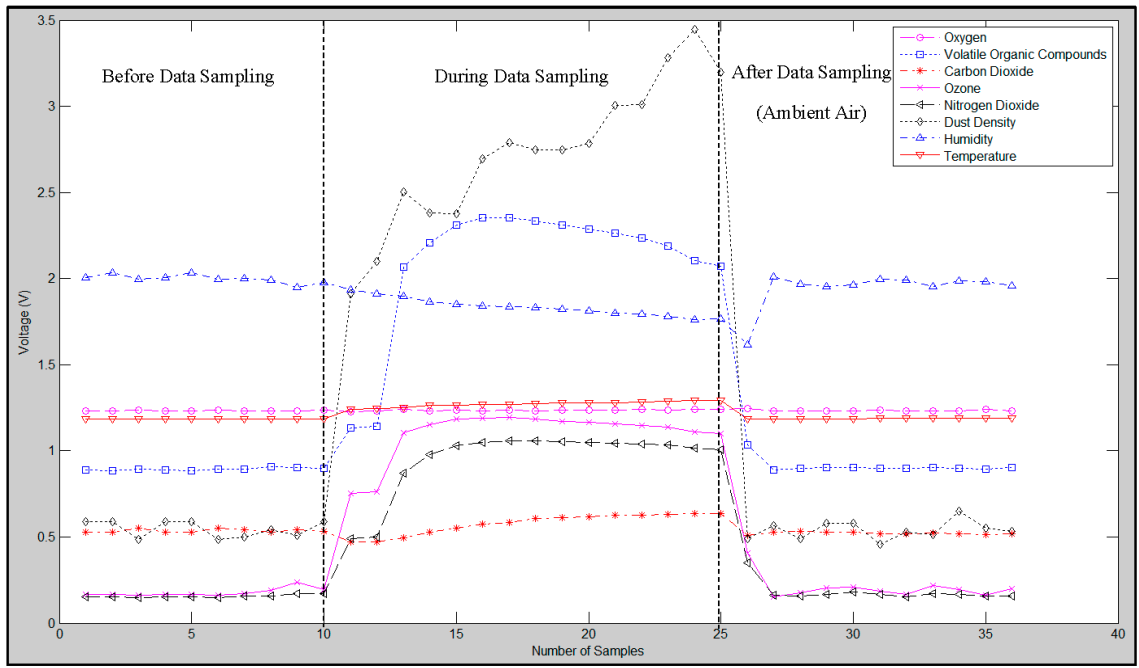

(**a**)

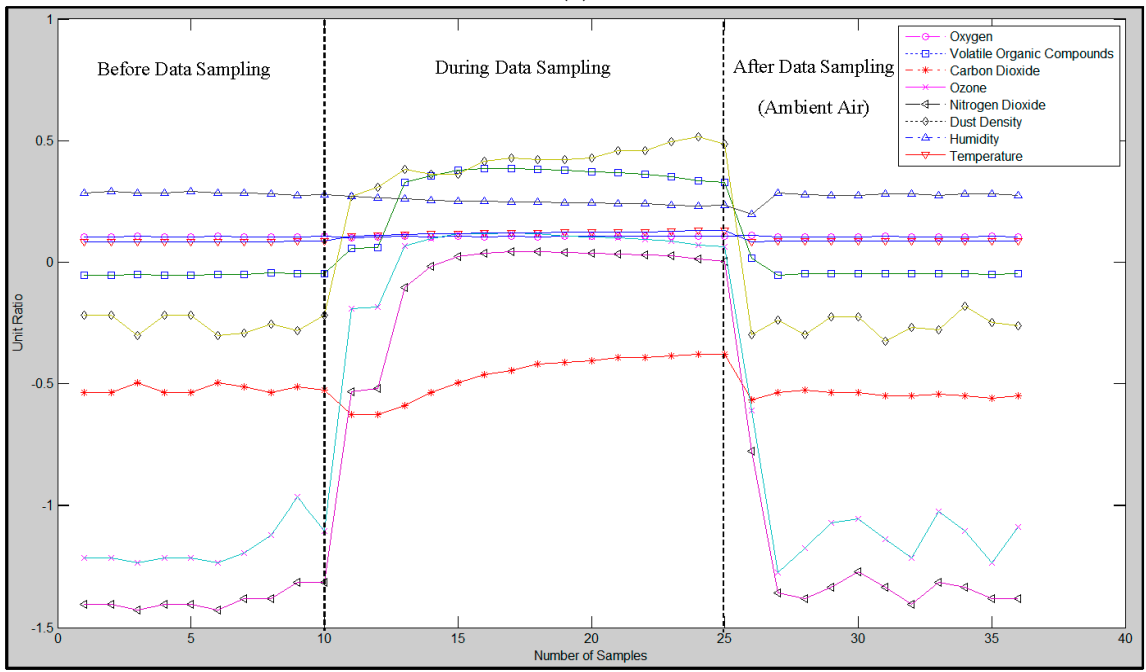

(**b**)

**Figure 2.** (**a**) Example of raw data for a scorching smell generated by paper at 250 °C; (**b**) The RLSSV feature extracted from the scorching smell of paper at 250 °C in (a).

### 2.4. Feature Selection

In this feature selection stage, the relative logarithmic sum squared voltage, the relative logarithmic voltage value, the relative sum squared voltage value, the relative voltage value, and the fractional voltage value, of the signal have been obtained. The selected features are chosen to investigate their performance on early fire data. The features have been tested for their reliability by examining the classification accuracy with a Probabilistic Neural Network (PNN). PNN and its function in this paper is explained further in Section 2.7. Out of the five features, the three best features with the highest classification accuracy are selected for dimensional reduction using PCA.

## 2.5. Dimension Reduction Using PCA

PCA is a linear technique which transforms a dataset from its original m-dimensional form into a new and compressed n-dimensional form where n < m. Dimension reduction has been implemented to investigate its effects on classification. Since the number of observations is reduced after the dataset is dimensionally reduced, the training period of PNN classifier will be minimized [16]. Thus, PCA is helpful not only in reducing the input variables of a dataset, but it also indirectly increases the classification ability of a classifier.

PCA gives the same number of principal components as the number of input variables. For example, if the data matrix has a dimension of 100 rows and 10 columns, the data matrix could be reduced to a 100 rows and three column matrix of principal components, without removing any important information from the original dataset. The data is arranged according to the variances between the classes, starting from highest variances descending from first column up to n numbered columns. However, out of the n reduced principal components, not all the principal components are needed to represent the data. Thus, the principal components need to be tested to find the appropriate number of principal components required for feature fusion. As explained in previous studies the optimal number of principal components can be obtained using a few criteria, such as the Broken stick model, Velicer's partial correlation procedure, cross-validation, Bartlett's test for equality of eigenvalues, Kaiser's criterion, Cattell's scree test and cumulative percentage of variance [17], which basically explais how much variances we are about to retain in the data. Based on this, in this study, eight principal components have been selected to observe the effect on the classification accuracy of PNN. For each selected feature in IAQ dataset, eight principal components have been obtained from eight input variables while for PEN3 dataset, 10 principal components have been obtained from 10 input variables. The latent, proportion and cumulative percentage corresponding to the principal component value from the principal components for the relative voltage value feature in the IAQ dataset and PEN3 dataset are given in Tables 2 and 3, respectively.

**Table 2.** Latent, proportion, and cumulative values of selected principal components for relative voltage value feature in the IAQ dataset.

| Principal Component | Latent | Proportion | Cumulative |
|---|---|---|---|
| 1 | 0.1064 | 0.4813 | 0.4813 |
| 2 | 0.0474 | 0.2141 | 0.6954 |
| 3 | 0.0335 | 0.1517 | 0.8471 |
| 4 | 0.0144 | 0.0650 | 0.9121 |
| 5 | 0.0096 | 0.0435 | 0.9556 |
| 6 | 0.0073 | 0.0329 | 0.9886 |
| 7 | 0.0019 | 0.0085 | 0.9970 |
| 8 | 0.0007 | 0.0030 | 1.0000 |

**Table 3.** Latent, proportion, and cumulative values of selected principal components for relative voltage value feature in the PEN3 dataset.

| Principal Component | Latent | Proportion | Cumulative |
|---|---|---|---|
| 1 | 7.8692 | 0.5338 | 0.5338 |
| 2 | 3.5164 | 0.2385 | 0.7723 |
| 3 | 1.8546 | 0.1258 | 0.8981 |
| 4 | 0.7612 | 0.0516 | 0.9497 |
| 5 | 0.4236 | 0.0287 | 0.9784 |
| 6 | 0.2476 | 0.0170 | 0.9954 |
| 7 | 0.0461 | 0.0030 | 0.9984 |
| 8 | 0.0176 | 0.0012 | 0.9996 |
| 9 | 0.0041 | 0.0003 | 0.9999 |
| 10 | 0.0015 | 0.0001 | 1.0000 |

## 2.6. Feature Fusion

In the feature fusion stage, the dimensionally reduced features have been fused to form the proposed IAQ-PCA hybrid feature for the IAQ dataset and the proposed PEN3- PCA hybrid feature for the PEN3 database. A similar approach was also reported by Luo who proposed an adaptive sensory fusion method for fire detection and isolation for intelligent building systems [18]. The proposed features have been tested and compared with the other normalised features mentioned in Section 2.3. The result of classification trials will be shown in Section 3. The feature fusion process for the IAQ-PCA hybrid features is shown in Figure 3. A similar process was also repeated for the PEN3-PCA hybrid features.

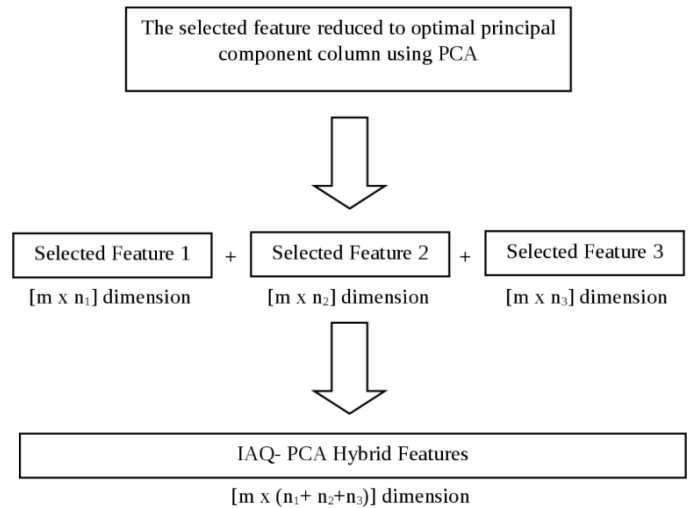

**Figure 3.** Feature Fusion Process for IAQ- PCA Hybrid Features.

## 2.7. Probabilistic Neural Network

Probabilistic Neural Network is highly regarded as a biologically inspired approach in classification as it functions similar to the human cognitive system. It requires less computational time and processing power compared to other classifiers. The human brain receives the input pattern from the nerves, compares it to the pattern in memory, and sums it together with other input patterns to find the probability that an event will occur [3]. Thus, in this work, PNN has been selected and used as a core classifier.

PNN can be used for classifying different input patterns. It was proposed by Specht based on Bayesian classification and the probability density function using classical estimators. Compared to the conventional multi-layer perceptron (MLP) classifier which uses a sigmoidal activation function, PNN uses an exponential activation function in its algorithm. The computational time for PNN is also much less than for the MLP classifier [3]. For example, let us consider a simple two class problem:

Classifying two classes problem, class A and class B.

The estimator for the probability density function as given in Equation (6) has been used in PNN:

$$f_A(X) = \frac{1}{(2\pi)^{n/2}} \frac{1}{m_A} \sum_{i=1}^{m_A} \exp\left[-\frac{(X - X_{Ai})^T (X - X_{Ai})}{2\sigma^2}\right] \qquad (6)$$

where, $X_{Ai}$ is the $i^{th}$ training pattern from class A, $n$ is the dimension of the input vectors, $m_A$ is the number of training patterns in class A, $T$ is the transpose of the value and $\sigma$ is a smoothing parameter corresponding to the standard deviation of the Gaussian distribution. This is the standard probability density function estimator used commonly in PNN and other neural networks. There are also some works highlighting on the modification in the exponential power of Equation (6), for example, normal, log- normal, Rayleigh and Weibull probability density functions which intend to provide better estimations of unknown stochastic processes, which do not require either an *a priori* choice of a mathematical model or the elaboration of the data histogram, but only the computation of the variability range of each components of available data samples [20].

Similar to our biological brain, the probabilistic neural network has four operational units known as input units, pattern units, summation units and output units. When PNN is given an input, the pattern unit will calculate the distance between the input vector and the trained input vectors. A vector with the information regarding the distance between the input and the training input is produced and passed to the summation unit. The contributions for each class of input are summed by the summation unit and a net output is generated. The net output has the information of the maximum of the probabilities to indicate a 1 for the specific class or a 0 for the other class.

The steps involved in the PNN algorithm are described below:

Step 0: Initialize the weights

Step 1: For each training input to be classified, do Step 2 to 4

Step 2: Pattern units:

Compute the net input to the pattern units:

$$Z_{inj} = x(w_j) = x^T w_j \tag{7}$$

Compute output Equation (8) using Equation (7):

$$Z_{outj} = \exp\left[\frac{z_{inj} - 1}{\sigma^2}\right] \tag{8}$$

Step 3: Summation unit:

Sum the inputs from the pattern units to which they are connected. The summation unit for class B multiplies its total input by Equation (9):

$$V_B = -\frac{P_B C_B m_A}{P_A C_A m_B} \tag{9}$$

Where:

$P_A$ & $P_B$ are the priori probability of occurrence of patterns in Class A and Class B,

$C_A$ & $C_B$ are the cost associated with classifying vectors in Class A and B, and

$m_A$ & $m_B$ are the number of training patterns in Class A and Class B.

Step 4: Output (decision) unit:

The output unit sums the signals from f$_A$ and f$_B$. The input vector is classified as Class A if the total input to the decision unit is positive. Based on the above example, the PNN network can classify two different classes when the input patterns of both classes are given to it. However, training the network with more sample inputs improves the ability of PNN. The degree of nonlinearity of the decision boundaries of PNN can be controlled by varying the spread factor, σ. Large values of σ make the decision boundary approach a hyperplane, while having a relatively small value approaching zero for σ gives a good approximation for highly nonlinear decision surfaces of PNN [3].

Consequently, in this paper, PNN is used to select the dominant features and to test the classification accuracy of the proposed and dominant features in distinguishing various materials involved in incipient fire cases. The PNN architecture is shown in Figure 4.

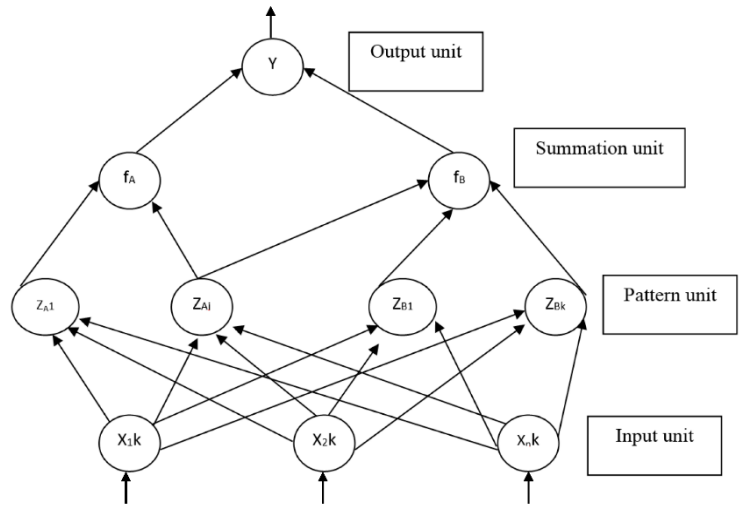

**Figure 4.** PNN Architecture;

The overall process flow of proposed multi- stage feature selection and fusion for both datasets is shown in Figure 5.

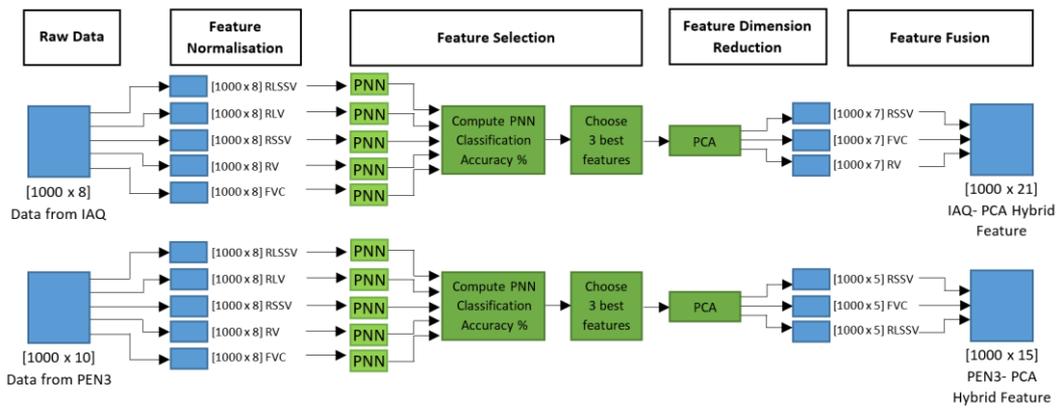

**Figure 5.** Multi-stage Feature Selection and Fusion Process Flow.

## 3. Results and Discussion

A Probabilistic Neural Network has been applied for classification of scorching smells generated from the different materials. In this application, both raw datasets have been subjected to the PNN classifier to select the most dominant features, prior to dimension reduction.

**Table 4.** PNN architectures.

| Parameters | Value for the IAQ Dataset | Value for PEN3 Dataset |
|---|---|---|
| Number of input neurons | 8 | 10 |
| Number of output neurons | 9 | 9 |
| Spread factor | 0.08 | 0.08 |
| Testing Tolerance | 0.001 | 0.001 |
| Number of training samples | 600 | 600 |
| Number of validation samples | 100 | 100 |
| Number of testing samples | 300 | 300 |
| Total number of samples | 1000 | 1000 |

The parameters used in PNN are shown in Table 4. As mentioned earlier in Section 2.7, the spread factor can be varied to control the degree of nonlinearity of the decision boundaries. It is the most important factor which influences the classification performance of the classifier. Therefore, the spread factor has been varied in these experiments to obtain the best classification performance [15]. The best value for spread factor for both datasets is recorded to be 0.08.

Classification performances have been computed for the nine classes for the IAQ dataset and PEN3 dataset as shown in Table 5. The classification accuracy of the each feature is clearly shown in the table. The classification result has been obtained by averaging the classification accuracy for 50 repetitions.

**Table 5.** Average PNN classification accuracies of features for IAQ and PEN3 datasets.

| Features | IAQ | | | PEN3 | | |
|---|---|---|---|---|---|---|
| | Minimum Classification Accuracy (%) | Maximum Classification Accuracy (%) | Average Classification Accuracy (%) | Minimum Classification Accuracy (%) | Maximum Classification Accuracy (%) | Average Classification Accuracy (%) |
| RLSSV | 97.11 | 99.41 | 98.75 | 97.15 | 99.54 | 99.29 |
| RLV | 97.64 | 98.65 | 98.31 | 97.43 | 99.02 | 98.84 |
| RSSV | 97.31 | 99.16 | 98.90 | 98.16 | 100.00 | 99.75 |
| RV | 97.36 | 99.43 | 98.81 | 98.19 | 99.45 | 99.12 |
| FVC | 97.42 | 99.14 | 98.84 | 98.41 | 99.55 | 99.51 |

For each dataset, the three best features with the highest classification accuracy have been selected for dimensional reduction with PCA. For the IAQ dataset, it is observed that RSSV, FVC and RV give the best accuracies, 98.90%, 98.84% and 98.81%, respectively. The PEN3 dataset, on the other hand, has RSSV, FVC and RLSSV with 99.75%, 99.51% and 99.29%, respectively, as its best features.

The three select features have eight columns each (inputs from eight gas and electrochemical sensors). At this stage, the dimension of each feature has been reduced to remove the redundant data and to select only the optimal number of features with high variance between classes, which is sufficient to represent the fire signature. Reducing the dimensions of the original data indirectly increases the classification accuracy and reduces the processing time of the classifier. The selection of principal component values in PCA will determine how much the dimensions of the m-dimension dataset will be reduced. The performance of the classifier has been investigated by varying the principal component values and the results have been recorded in Table 6.

**Table 6.** Average PNN classification results in % for selecting principal component values in PCA for the IAQ and PEN3 datasets.

| Principal Component Value | IAQ | | | PEN3 | | |
|---|---|---|---|---|---|---|
| | RSSV | FVC | RV | RSSV | FVC | RLSSV |
| 1 | 74.07 | 75.30 | 74.47 | 83.26 | 82.58 | 82.12 |
| 2 | 82.43 | 83.11 | 83.56 | 87.51 | 87.39 | 87.03 |
| 3 | 87.74 | 87.27 | 88.28 | 91.97 | 91.67 | 90.97 |
| 4 | 90.17 | 90.21 | 90.21 | 98.28 | 97.95 | 97.49 |
| 5 | 95.62 | 95.66 | 95.45 | 100.00 | 99.91 | 99.76 |
| 6 | 98.30 | 98.13 | 97.70 | 98.75 | 98.66 | 98.12 |
| 7 | 99.02 | 99.02 | 98.96 | 97.35 | 97.12 | 96.81 |
| 8 | 98.88 | 98.80 | 98.86 | 96.74 | 96.55 | 96.26 |

As seen in Table 6, 6–8 principal components give the most successful classification results for the IAQ dataset, while 4–6 principal components give the most successful classification results for the PEN3 dataset. The range of classification accuracies range from a minimum of 98.13% to a maximum 99.02% for the IAQ dataset, and from a minimum of 97.49% to maximum of 100.00% for the PEN3 dataset. Out of this range, the best classification accuracies for the IAQ dataset have been observed to occur when the principal component value is seven, while, for the PEN3 dataset, the optimal principal component value has been observed to be five. Thus, the dimensions of the IAQ and PEN3 datasets have been reduced to seven and five principal components scores, respectively. The dimensionally reduced features have been fused to form the proposed IAQ-PCA hybrid feature for the IAQ dataset and the proposed PEN3-PCA hybrid feature for the PEN3 database. The fused feature for the IAQ dataset is a matrix of 1000 rows and 21 columns, while the fused feature for the PEN3 dataset is a matrix of 1000 rows and 15 columns.

The confusion matrixes of PNN of both the IAQ-PCA hybrid feature and the PEN3-PCA hybrid feature for classification trials and its respective mean classification accuracy for 50 repetitions have been tabulated in Tables 7 and 8. Both tables consist of the true positive, true negative, false positive and false negative counts, which are useful in computing performance evaluation of the PNN classifier. M1 denotes material 1, and NA denotes normal air.

**Table 7.** Confusion Matrix of PNN of proposed IAQ-PCA hybrid feature for 50 repetitions.

| | | Actual | | | | | | | | | |
|---|---|---|---|---|---|---|---|---|---|---|---|
| | | **M1** | **M2** | **M3** | **M4** | **M5** | **M6** | **M7** | **M8** | **NA** | **Mean Classification Accuracy (%)** |
| Predicted | **M1** | 40 | 0 | 0 | 0 | 0 | 0 | 0 | 0 | 0 | 100.00 |
| | **M2** | 0 | 39 | 0 | 0 | 0 | 1 | 0 | 0 | 0 | 99.52 |
| | **M3** | 0 | 0 | 40 | 0 | 0 | 0 | 0 | 0 | 0 | 100.00 |
| | **M4** | 0 | 0 | 1 | 39 | 0 | 0 | 0 | 0 | 0 | 99.12 |
| | **M5** | 0 | 0 | 0 | 0 | 39 | 0 | 1 | 0 | 0 | 99.01 |
| | **M6** | 1 | 0 | 0 | 0 | 0 | 39 | 0 | 0 | 0 | 99.51 |
| | **M7** | 0 | 0 | 0 | 0 | 0 | 0 | 40 | 0 | 0 | 100.00 |
| | **M8** | 0 | 0 | 0 | 0 | 0 | 0 | 1 | 39 | 0 | 99.15 |
| | **NA** | 0 | 0 | 0 | 2 | 0 | 0 | 0 | 0 | 78 | 99.24 |

**Table 8.** Confusion Matrix of PNN of proposed PEN3-PCA hybrid feature for 50 repetition.

| | | Actual | | | | | | | | | |
|---|---|---|---|---|---|---|---|---|---|---|---|
| | | **M1** | **M2** | **M3** | **M4** | **M5** | **M6** | **M7** | **M8** | **NA** | **Mean Classification Accuracy (%)** |
| Predicted | **M1** | 40 | 0 | 0 | 0 | 0 | 0 | 0 | 0 | 0 | 100.00 |
| | **M2** | 0 | 40 | 0 | 0 | 0 | 0 | 0 | 0 | 0 | 100.00 |
| | **M3** | 0 | 0 | 40 | 0 | 0 | 0 | 0 | 0 | 0 | 100.00 |
| | **M4** | 0 | 0 | 0 | 40 | 0 | 0 | 0 | 0 | 0 | 100.00 |
| | **M5** | 0 | 0 | 0 | 0 | 40 | 0 | 0 | 0 | 0 | 100.00 |
| | **M6** | 0 | 0 | 0 | 0 | 0 | 40 | 0 | 0 | 0 | 100.00 |
| | **M7** | 0 | 0 | 0 | 0 | 0 | 0 | 40 | 0 | 0 | 100.00 |
| | **M8** | 0 | 0 | 0 | 0 | 0 | 0 | 0 | 40 | 0 | 100.00 |
| | **NA** | 0 | 0 | 0 | 0 | 0 | 0 | 0 | 0 | 80 | 100.00 |

The performance evaluation of a classifier can be performed by examining a few statistical measures obtained by calculating the sensitivity, specificity and accuracy scores for the classifier [19]. The sensitivity is the division of the correctly selected decisions over the total decisions which are actually the deserved selections, as shown in Equation (10). The specificity (Equation (11)) indicates the division of correctly rejected decisions by the total decisions which actually deserve rejection. The accuracy is the score of correctly decided decisions over the total decisions made. The accuracy formula is shown in Equation (12):

$$\text{Sensitivity} = \frac{TP}{TP+FN} \times 100\% \tag{10}$$

$$\text{Specificity} = \frac{TN}{TN + FP} \times 100\%, \text{ and} \tag{11}$$

$$\text{Accuracy} = \frac{TP + TN}{TP + TN + FP + FN} \times 100\% \tag{12}$$

where, the TP indicates the true positive decisions, FP is the false positive decisions, TN is the true negative decisions and FN is the false negative decisions. Based on Table 7, TP is 315, FP is 5, TN is 78 and FN is 2.

Both hybrid features have been compared with the other best features selected as discussed earlier through Table 5 for both the IAQ and PEN3 datasets. Tables 9 and 10 show that the proposed IAQ-PCA and PEN3-PCA hybrid features have better performances compared to the standard normalised features. The IAQ-PCA hybrid feature recorded a highest accuracy value of 98.25%, while the PEN3-PCA hybrid feature recorded a highest accuracy of 100%.

**Table 9.** Average PNN classification results comparison between the best features for the IAQ dataset.

| Feature | Sensitivity (%) | Specificity (%) | Accuracy (%) |
|---|---|---|---|
| IAQ-PCA Hybrid Feature | 99.37 | 93.98 | 98.25 |
| RSSV | 99.05 | 91.67 | 97.50 |
| FVC | 98.74 | 91.57 | 97.25 |
| RV | 99.04 | 89.53 | 97.00 |

**Table 10.** Average PNN classification results comparison between the best features for the PEN3 dataset.

| Feature | Sensitivity (%) | Specificity (%) | Accuracy (%) |
|---|---|---|---|
| PEN3-PCA Hybrid Feature | 100.00 | 100.00 | 100.00 |
| RSSV | 99.85 | 96.17 | 99.75 |
| FVC | 99.63 | 96.85 | 99.51 |
| RLSSV | 99.51 | 96.09 | 99.29 |

The proposed features have been compared with other common available classifiers. Feed-forward Neural Network (FFNN), Elman Neural Network (ENN) and k- Nearest Neighbour (kNN) classifiers have been selected for this purpose. The comparison results between the classifiers for the proposed PCA-based hybrid features are presented in Table 11.

**Table 11.** Average classification results comparison between different classifiers for proposed PCA based hybrid features.

| Classifier | IAQ | | | PEN3 | | |
|---|---|---|---|---|---|---|
| | Sensitivity (%) | Specificity (%) | Accuracy (%) | Sensitivity (%) | Specificity (%) | Accuracy (%) |
| PNN | 99.75 | 92.63 | 98.25 | 100.00 | 100.00 | 100.00 |
| FFNN | 98.71 | 91.53 | 97.16 | 99.88 | 95.47 | 99.75 |
| ENN | 98.53 | 91.64 | 97.65 | 99.78 | 94.57 | 99.74 |
| kNN | 99.41 | 91.42 | 97.89 | 99.89 | 95.91 | 99.85 |

For FFNN and ENN, the number of hidden layers, the learning rate, the momentum factor, and the type of activation functions have been modified to obtain the best classification performance. The architectures of the classifiers have been modelled to have 21 input neurons, 45 hidden neurons and nine output neurons for the IAQ-PCA hybrid feature, and 15 input neurons, 32 hidden neurons and nine output neurons for the PEN3-PCA hybrid feature, respectively. The learning rate has been set at 0.001 and the momentum factor is 0.85 for both classifiers. In addition, the activation function, the testing tolerance and the maximum iteration have been tuned to log-sigmoid, 0.00001 and 1000, respectively. The backpropagation algorithm has been utilised for the weights training. For the kNN classifier, the $k$ value has been set to 3 for the IAQ-PCA feature. For the PEN3-PCA feature, the $k$ value is set at 1. The $k$ value in the kNN classifier is extremely training data dependent. Having cross-validation methods such as K- fold and leave-one-out are useful to find the $k$ value which leads to the highest classification generalizability. In these paper, all the parameters involved in these classifiers have been selected based on trial and error to get the best classification accuracy. As seen on Table 11, the sensitivity, specificity and accuracy of each classifier have been tabulated for both features. From the table, it can be clearly seen that the dimensional reduction and fusion of the features to form hybrid features has deliberately increased the classification accuracy of the classifiers. The success rate of PCA-based hybrid features in the PNN classifier surpasses the performance of other common classifiers.

## 4. Conclusions

Feature selection and feature reduction have been demonstrated in detail. Both combined features from IAQ and PEN3 gives better classification accuracy. In this paper, a PCA-PNN-based feature selection technique has been proposed and investigated. The data has gone through various stages of processing such as normalised feature extraction, feature verification, binary data normalisation, PCA and data randomisation, before it is fed to the classifier. For investigation purposes, PNN has been selected as the classifier and the results have been further tested using other classifiers on the two datasets, The IAQ dataset from the in-house system and the PEN3 dataset from a commercial electronic nose system. As a result, the PEN3 dataset has better classification performance compared to the IAQ dataset for all the comparisons. This could be due to the sensitivity of the PEN3 electronic nose's gas sensors and the data capturing ability of the Winmuster software, which is used commercially. It is also observed from the analysis that the performance of the IAQ electronic nose is almost comparable to that of the PEN3 electronic nose. Thus, it is proven to be useful for early fire detection and prediction of various incipient stage scorching materials.

**Acknowledgments:** This research work is supported by Malaysian Technical University Network (MTUN) COE research grant (grant number: 9016-00010), Ministry of Education Malaysia under the *Skim Latihan Tenaga Pengajar Akademik IPTA* (SLAI) scholarship and Centre of Excellence for Advanced Sensor Technology (CEASTech), Universiti Malaysia Perlis, Malaysia.

**Author Contributions:** All authors have agreed with the design of experiment for the research which is prepared according to the European Standard. Ammar Zakaria and Ali Yeon Md Shakaff have contributed with the correction and critical comments on the manuscript. Shaharil Mad Saad has designed the low cost electronic nose (IAQ) unit incorporating gas and electrochemical sensors. Ali Yeon Md Shakaff has given permission to use the PEN3 unit for the experiments.

**Conflicts of Interest:** The authors declare no conflict of interest.